\def\year{2022}\relax
\documentclass[letterpaper]{article} 
\pdfoutput=1
\usepackage{aaai21}  
\usepackage{times}  
\usepackage{helvet} 
\usepackage{courier}  
\usepackage[hyphens]{url}  
\usepackage{graphicx} 
\usepackage{natbib}  
\usepackage{caption} 
\usepackage{comment}
\title{Information Operations in Turkey:\\Manufacturing Resilience with Free Twitter Accounts}
\author{
    Maya Merhi\hspace{0.5in} Sarah Rajtmajer\hspace{0.5in} Dongwon Lee
}

\affiliations{
\\ 
    The Pennsylvania State University, USA\\

    \texttt{\{mvm6917, smr48, dongwon\}@psu.edu} 

}

\begin{document}

\maketitle

\begin{abstract}
Following the 2016 US elections Twitter launched their Information Operations (IO) hub where they archive account activity connected to state linked information operations. In June 2020, Twitter took down and released a set of accounts linked to Turkey's ruling political party (AKP). We investigate these accounts in the aftermath of the takedown to explore whether AKP-linked operations are ongoing and to understand the strategies they use to remain resilient to disruption. We collect live accounts that appear to be part of the same network, ~30\% of which have been suspended by Twitter since our collection. We create a BERT-based classifier that shows similarity between these two networks, develop a taxonomy to categorize these accounts, find direct sequel accounts between the Turkish takedown and the live accounts, and find evidence that Turkish IO actors deliberately construct their network to withstand large-scale shutdown by utilizing explicit and implicit signals of coordination. We compare our findings from the Turkish operation to Russian and Chinese IO on Twitter and find that Turkey's IO utilizes a unique group structure to remain resilient. Our work highlights the fundamental imbalance between IO actors quickly and easily creating free accounts and the social media platforms spending significant resources on detection and removal, and contributes novel findings about Turkish IO on Twitter.
\end{abstract}

\section{Introduction}

Social networking sites have become the birthplaces of conspiracy theories, battlegrounds for trolling, and the new stage for \textit{information operations} (IO), which describes ``actions taken by governments or organized non-state actors to manipulate public opinion'' ~\cite{arif_acting_2018}. The most widely studied of such has been the Russian Internet Research Agency's (IRA) campaign aimed at influencing the outcome of the 2016 U.S. Presidential Election. As authorities began investigating this effort in the aftermath of Trump's victory in 2016, social media corporations were urged to investigate and report on the manipulation that took place ~\cite{starbird_disinformation_2019}. In their response, Twitter began releasing archives of state-linked IO as they detect and remove them. There are currently archives from 20 different countries that include account information, tweets, and embedded media files ~\cite{twitter}. 

Amongst these is a large dataset representing an operation Twitter has attributed to the Turkish Justice and Development Party (Adalet ve Kalkınma Partisi; AKP) \cite{turkey_announcement_2020}. Turkey has been undergoing a hostile political transformation into an authoritarian regime for the past decade, which became even more severe following a failed coup attempt in 2016 ~\cite{yilmaz_turkeys_2019}. AKP has a well-known history of utilizing social media to spread pro-government content, and utilizing an army of troll accounts commonly known as the AK Trolls to attack those critical of the government ~\cite{saka_social_2018, bulut_mediatized_2017,albayrak_turkeys_2013}. In 2020, Twitter 
removed 7,340 accounts linked to the AKP from the platform and subsequently shared these accounts and their activity with researchers.

Using these accounts as a starting point, we  explore the following research questions in this work: (RQ1) 
{\em 
Is there evidence that AKP-linked operations are ongoing following Twitter's takedown of 7,340 AKP-linked accounts in 2020? }
(RQ2) {\em What strategies do ongoing AKP-linked operations use to remain resilient to disruption?} To address these questions, we collect live accounts that 
appear to be associated with the operation, including 169 accounts which we suggest have been stood up as direct replacements for accounts suspended in Twitter's takedown. We find that a rich interaction network underlies the operation and remains robust to node removal over time.  
Our findings suggest that Turkish IO actors prepare for and adapt to shutdown by creating a strong network through explicit signals of coordination such as hashtags and following patterns, and strategic creation and assignment of new accounts. We compare the Turkish network with known Russian and Chinese IO and find that all three networks exhibit explicit signals of coordination, but the Turkish network utilizes a unique group structure to build resilience. These findings lay groundwork for a broader understanding of how state-linked IO adapt in the face of detection.


\section{Related Work}


\textbf{Information Operations.} 
Social media has inadvertently created direct channels for IO actors to disseminate targeted propaganda to a wide audience. 
There is evidence of social media manipulation campaigns originating from authoritarian regimes as well as democratic states and targeting both domestic and foreign audiences \cite{bulut_mediatized_2017,Ferrara_2017,farkas_ira_2018}. Authoritarian regimes primarily target domestic audiences through carefully crafted narratives disseminated by fake accounts and legitimate accounts of local government officials \cite{king2017chinese}. \citet{bradshaw_troops_2017} found that of 28 countries examined, every authoritarian regime used social media campaigns to target their domestic populations, while almost all democratic nations in their study organize campaigns to target foreign audiences.

\noindent \textbf{IO Strategy and Tactics.} 
Over the past decade, IO have included coordinated efforts within the social media environment in support of strategic aims ~\cite{lin_cyber-enabled_2019,starbird_disinformation_2019,bradshaw_troops_2017}. These efforts employ fabricated profiles, intricately constructed narratives, and armies of social bots that amplify disinformation, e.g., ``troll armies'' ~\cite{arif_acting_2018,Ferrara_2017,recuero_hyperpartisanship_2020,bradshaw_troops_2017,linvill_troll_2020}.  

IO leverage content-based, language-based, cross-platform and multimedia strategies. Content-based strategies include organizing dedicated accounts to post certain content, synchronizing content with real-world events, and fabricating profiles that imitate real people to embed polarizing content in different communities \cite{farkas_ira_2018,zannettou_disinformation_2019,arif_acting_2018}. Language-based strategies include the targeting of populations using their native language (i.e., Russian troll accounts tweeting in German~\cite{zannettou_disinformation_2019}), and the use of deceptive language and shorter, less-complex language than baseline online users~\cite{addawood_linguistic_2019}. Cross-platform and multimedia strategies include posting links to external websites, posting links to accounts on other social media platforms, and using photos and videos to increase content sharability ~\cite{zannettou_characterizing_2020,zannettou_who_2019}.


\noindent \textbf{Detecting IO.} 
Few works have used Twitter's archives of detected IO to classify and detect live IO. \citet{alizadeh_content-based_2020} utilize Twitter's IO archives originating from Russian, China, and Venezuela to build content-based Random Forest classifiers for a number of classification tasks, including using a prior data release to classify trolls in a subsequent data release. 
\citet{luceri_detecting_2020} employ Inverse Reinforcement Learning (IRL) to identify Russian troll accounts leading up to the 2016 US Presidential Election. They utilize IRL to infer users' objectives and identify how trolls' incentives differ from regular online users. \citet{im_still_2020} utilize Twitter's first release of ground-truth Russian troll data to build machine learning classifiers of English-speaking Russian trolls targeting US politics. 

\noindent \textbf{Turkey: The AK Trolls.} 
In 2013, the Gezi park protests mobilized Turkish citizens to Twitter as a primary news source and mechanism for political discourse and organization due to severe censorship in mainstream media \citep{yilmaz_turkeys_2019,karatas_online_2017}. In response, the AKP funded, recruited, and trained an army of 6000 young AKP members to create and disseminate pro-government/AKP content on social media \citep{albayrak_turkeys_2013,saka_social_2018}.
These ``AK Trolls''  use social media to spread AKP ideals through large volumes of messages and images \citep{saka_social_2018,albayrak_turkeys_2013}. 
They promote unwavering support and praise of President Erdoğan, continuously criticize and demonize opposing political parties, specifically the People's Republican Party (CHP) and the People's Democratic Party (HDP), and attack journalists who criticize the government \citep{sio1,bulut_mediatized_2017,saka_social_2018,karatas_online_2017,basaran_digital_nodate}. Twitter's takedown attributed to the AKP shares these same sentiments and is likely part of the AK Trolls' larger network of accounts  \citep{sio1}. Stanford Internet Observatory released a report on the Turkish takedown and identified tactics such as the use of compromise accounts, highly centralized retweet rings, time-coordinated posting, and coordinated content \citep{sio1}. \citet{karatas_online_2017} utilized digital ethnography to examine how the AKP utilizes political trolling strategies, how trolling is coordinated to silence opposing view points, and the effects of AK Trolls on vulnerable and unprotected citizens, e.g., self-censorship and quitting social media. \citet{saka_social_2018} provides interviews with self-proclaimed AK Trolls; the author finds that political trolling in Turkey ``is more decentralized and less institutionalized then generally thought'' \cite[p.1]{saka_social_2018}. \citet{bulut_mediatized_2017} provide an analysis of AK Trolls through the lens of mediated populism and demonstrate the increased politicization of Twitter in Turkey following the Gezi Park protests. 

Our work contributes a taxonomy of Turkish IO account types and groups memberships, a novel analysis of explicit signals used by IO networks, and evidence that Turkish accounts deliberately construct their network to withstand large scale shutdown.

\section{Dataset}


We analyze two datasets and their interactions: first, the set of accounts and content identified by Twitter as a Turkish IO and shared on the company's IO Hub\footnote{\url{https://about.twitter.com/en/our-priorities/civic-integrity\#data}} (here forward, \emph{takedown dataset}); second, a network of Turkish accounts that we detected live on Twitter as suspected Turkish IO actors (here forward, \emph{live dataset}) (see Table \ref{tab:datasets}).

 \begin{table}[h!]
 \small
  \begin{tabular}{|l|p{0.3\linewidth}|l|}
  \hline
   \textbf{Feature} & \textbf{Takedown} & \textbf{Live} \\ \hline
  \# of users &  7340 & 7973 \\ \hline
  \# of active users &  6270 & 7502 \\ \hline
  \# of tweets &  36,948,536 & 9,298,325 \\ \hline
   start date & 2009-06-04 & 2020-01-01 \\ \hline
  end date & 2020-04-21 & 2021-01-02 \\ \hline
  
 \end{tabular}
 \caption{\label{tab:datasets} Overview of takedown and live accounts. Number of active users includes all users who tweeted at least once.}

\end{table}

\subsection{Twitter's Turkish Takedown}
In June 2020, Twitter added 7340 accounts and nearly 37 million tweets attributed to Turkey's AKP to their archive.\footnote{\url{https://blog.twitter.com/en_us/topics/company/2020/information-operations-june-2020.html}} The downloadable data hashes the User ID, username, and display name of all users with less than 5000 followers. We use the unhashed data, which researchers can apply to access. 
Throughout the paper, we suppress any usernames that are hashed in the public version of the dataset. 



\subsection{Live Account Collection}
We collect live accounts and tweets suspected to be part of the Turkish IO through analysis of the social networks and common behaviors amongst accounts in Twitter's takedown dataset. We start with the follow trains and groups observed in the takedown dataset \cite{sio1}. 
Follow trains are tweets that mention (@) a list of users and request their followers follow these accounts. Of almost 37 million tweets released in Twitter's takedown dataset, approximately 1.1 million are follow trains.\footnote{We calculate this by filtering tweets that contain at least 5 mentions and have a mention to word ratio above 0.8.} Many Turkish IO accounts are also members of ``groups'' and declare their group membership in their profile description.  These accounts create ``RT'' (retweet) and ``fav'' (favorite) groups to share content and boost followers, engagement, and visibility. 

(1) We manually identify 13 ``parent'' accounts that appear to be part of the operation based on their interactions with and strong similarities to accounts from the takedown. Specifically, these accounts frequently retweeted and mentioned multiple takedown accounts, and likewise were retweeted and mentioned by takedown accounts. They also posted very similar content to accounts in the takedown (similar political images/messages, follow trains, targeted hashtag campaigns). And, they contained other elements common to Turkish IO, e.g., Turkish flag imagery, profile descriptions devoted to the AKP (\#AKParti, \#RecepTayyipErdoğan), roughly equal followers to friends ratio.\footnote{All of the user IDs and tweet IDs we collected are available to the community on Github \emph{https://github.com/mayamerhi/Turkey\_IO\_ICWSM23}. Parent accounts are denoted as such.}

(2) We collect parent accounts' followers and friends through Twitter API's GET friends/list and GET followers/list endpoints. To 
reduce potential false positives, we remove Recep Tayyip Erdoğan's account and all accounts he was following at the time-- a total of 96 accounts, almost all of which were verified, and all appeared to be real Turkish politicians or organizations. 

(3) As our focus was to study operations in response to the takedown, we filter out accounts created before 2020, resulting in 7973 accounts. Twitter announced the takedown on June 11, 2020. Based on date of last tweet, it appears that most accounts were suspended prior to Jan 2020.  

(4) To collect tweets from live accounts, we use the Twitter API GET statuses/user\_timeline endpoint, which returns up to 3200 of a user's most recent Tweets. 
Our tweet collection resulted in 7973 accounts and 9,298,325 associated tweets throughout 2020.

\subsection{Suspended Accounts}
Since our initial data collection in Dec 2020, a significant portion of the accounts we turned up have been suspended. Specifically, as of Dec 6 2021, 2454 ($\sim$31\%) of the accounts in our live dataset been suspended by Twitter.\footnote{To determine which, we queried the Twitter API for all users in our live dataset and assigned the label `suspended' for all accounts whose API response was \textit{account suspended}.}  
Twitter lists the three most common reasons for account suspension as: spam/fake account, account security at risk (hacked or compromised), and abusive tweets or behavior (threats or impersonating other accounts).\footnote{\url{https://help.twitter.com/en/managing-your-account/suspended-twitter-accounts}} As we collected these accounts based on interactions and shared characteristics with the takedown dataset, and as IO routinely 
engage in these behaviors, we suggest that accounts from our live dataset which have been suspended can be reasonably assumed IO accounts. We denote this subset of 2454 suspended accounts as \emph{live suspended}. 




\section{RQ1: Ongoing Operations}

We explore the takedown and live datasets to explore whether the live accounts are part of an ongoing AKP operation.
First, we use network analyses to map interactions between users from the takedown and live datasets. Then, we create a BERT-based classifier to distinguish Turkish state linked accounts from ``ordinary'' Turkish Twitter users and provide further evidence that suspended users in the live dataset are likely Turkish state linked users. 

\subsection{Network Interactions}
\label{sec_rq1}

\begin{figure}[h!]
\includegraphics[scale=0.3]{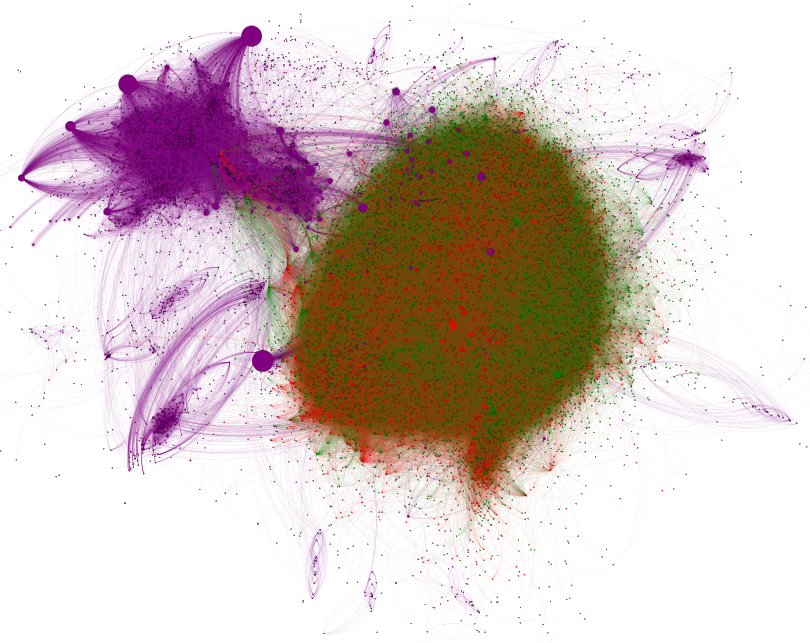}
\caption{\textbf{Account interaction graph.} Takedown accounts are represented in purple, live accounts (still live as of December 2021) are green, and suspended accounts are red. Edge color represents the color of the target node. If there are bidirectional interactions between users, both edges are present. Edges are weighted by number of interactions.} 
\label{fig:all_interaction_graph}
\end{figure}

Figure \ref{fig:all_interaction_graph} shows all interactions (mentions, retweets, replies, quotes) between users in both the live and takedown datasets. This includes interactions both within and across datasets. The graph contains a total of 11,551 user nodes and 609,459 directed, weighted interaction edges. The network shows that there are considerably more interactions amongst accounts in each dataset than between them. This is largely due to our collection. Namely, tweets in the takedown dataset span 2009 to early 2020 while the live accounts we pulled were created in 2020, so the two sets of accounts' activity overlapped only briefly. Specifically, only 1426 users in Twitter's takedown dataset were active in 2020, and 1422 tweeted their last tweet in Jan 2020. Therefore, most of the overlap between these two datasets happened in only one month (Jan 2020).

\begin{figure}[tb]
\includegraphics[scale=0.20]{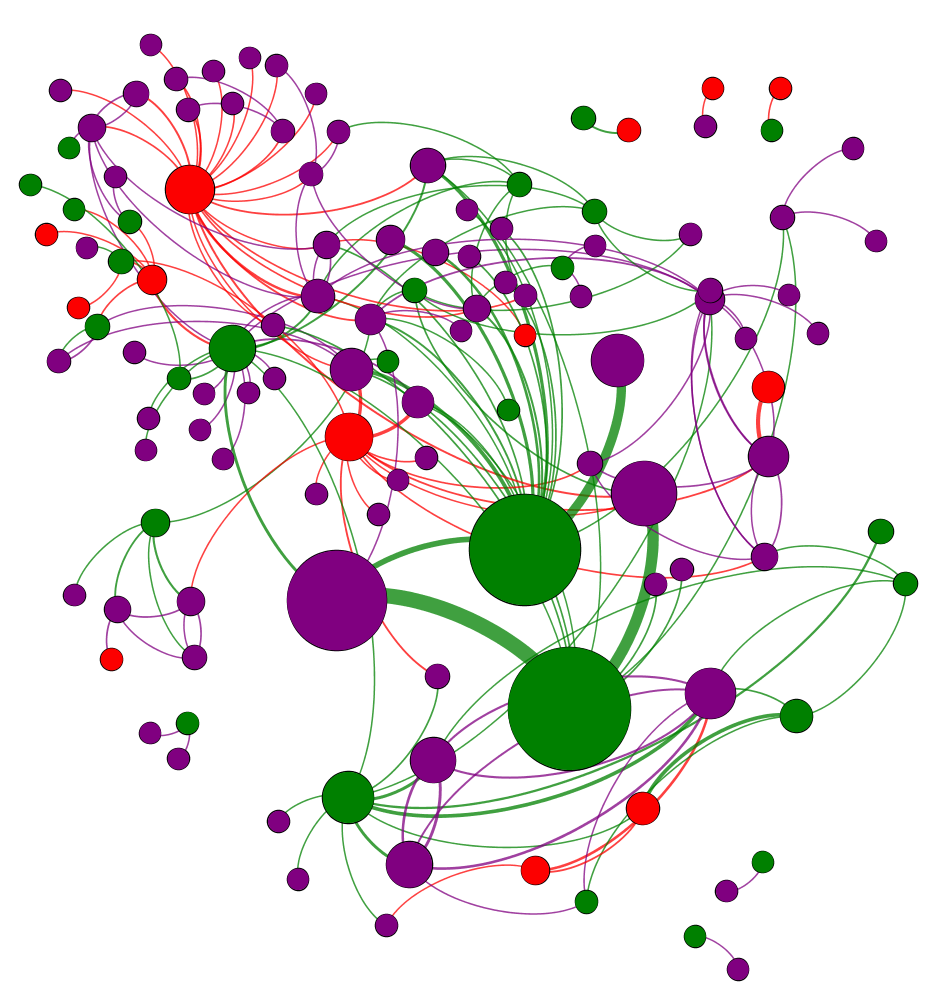}
\caption{\textbf{Transition network.} Interaction graph amongst all users in both the takedown and the live datasets during Jan 2020. Node/edge colors follow as in Figure \ref{fig:all_interaction_graph}.}
\label{fig:Jan20_overlap}
\end{figure}

In Figure \ref{fig:Jan20_overlap}, we zoom in on this time period to analyze the interactions amongst takedown and live accounts while they were both active. 
The graph contains 120 nodes and 192 weighted interaction edges. There are 80 nodes from the takedown dataset and 40 nodes from the live dataset, 13 of which have been suspended by Twitter as of Dec 6, 2021. Unlike Figure \ref{fig:all_interaction_graph}, we do not see two distinct clusters separating the two datasets. Rather, we see a high amount of interconnectivity between the two datasets. This view, we suggest, captures a snapshot of the operation's activity transition to new accounts in response ongoing shutdown.  

Of 1426 users in the takedown dataset active in 2020, 1422 last tweeted in Jan 2020.\footnote{Of the remaining users in the takedown set, 2143 were suspended in 2019 and 1239 were suspended in 2018.} This indicates two possibilities: Twitter may have suspended many accounts from the takedown dataset at once in Jan 2020, or the Turkish operation was already transitioning to new accounts in anticipation of shutdown. 
During this time, they began interacting with accounts in the live dataset to boost visibility of newly created accounts through retweets and follow trains. During Jan 2020, takedown accounts interacted with accounts from the live dataset 530 times.\footnote{This is greater than the number of edges in Figure \ref{fig:Jan20_overlap} because the graph contains weighted edges.} Of these 530 tweets, 455 were retweets and 72 were follow trains. 

\subsection{BERT Classification }
We create a BERT-based classifier trained on users from the takedown and evaluated on the live suspended users. We aim to determine whether latent similarities between the two datasets can further substantiate the link between them. We use the pre-trained Turkish language DistilBERTurk model \cite{stefan_schweter_2020_3770924}. We fine-tune DistilBERTurk using accounts from the takedown as positively labeled samples and we collect negatively labeled samples using the Twitter API. Our classifier yields an F1-score of 0.88 when tested on the takedown dataset, and 0.71 when evaluating the live dataset. 

\noindent \textbf{Classifier data.} Our positively labeled training dataset contains all tweets associated with the 3569 accounts in the takedown that posted in 2019 and 2020. 
We collect negatively labeled data to represent ordinary Turkish language Twitter users that have no link to the Turkish state. 
For a pseudo-random sample of such users, we queried the Twitter API for a list of common Turkish stopwords, while also restricting the search to return tweets written in the Turkish language, between the date ranges of Jan 1 2019 to Jan 1 2021. We collected the timelines for 6571 users. We use 3571 of these users as negatively labeled samples for training, validation, and testing (here after, \textit{negative dataset}) and 3000 as negatively labeled samples when evaluating the fine-tuned classifier on the live dataset. Since the takedown dataset provides \textit{all} tweets ever posted by users, and the Twitter API provides only up to 3200 tweets for the live dataset and negative data, we limit the number of tweets for each user to a random sample 2000 of their tweets. 

\noindent \textbf{Fine-tuning specifications.} We use 3569 users from the takedown dataset as positively labeled samples and 3571 users from the negative dataset as negatively labeled samples to create a balanced dataset of 7140 users for model building. We use a 70-15-15 training, validation, and testing split. We remove only URL links for text pre-processing. We use the auto-tokenizer from the DistilBERTurk model, the Keras implementation of the Adam optimizer with a learning rate of $5e^{-5}$, binary cross-entropy loss function, batch size of 16, and 2 training epochs. After training, validation, and testing, we run the fine-tuned model on the live suspended dataset to evaluate if the model trained on takedown users distinguishes live suspended users from negative users. 
Table \ref{tab:model_results} gives the model's performance.

\noindent \textbf{BERT results.} Our fine-tuned DistilBERTurk model performs well on the takedown users during testing and shows 
reasonable discernment between these users and the negative users. The model also performs well when evaluated on the live suspended users. These model results provide further evidence that users in the live dataset are likely Turkish state-linked users. The classifier correctly classifies 79\% of the live suspended accounts in the positive class, i.e., as Turkish state linked accounts rather than the ordinary Turkish Twitter users.\footnote{Of our 7973 collected users, 7502 of were active at the time of collection. Of the 2454 suspended users, 2093 tweeted at least once. We removed all users with no tweets from the classification task. Our classifier predicts that 1648 of the 2093 suspended accounts we predict on are in the positive class.} Of the remaining users in the live dataset, the classifier predicts 4181 users in the positive class and 1228 users in the negative class. 
Of these 1228 users that have not been suspended and which our classifier predicts to be ordinary users, it is likely that some subset are actual AKP followers. See Discussion and Conclusions for further discussion on actual AKP followers. 

As detailed in Table \ref{tab:model_results}, there is a slight decrease in classifier performance during evaluation on the live dataset, although we note that 0.71 F1-score is in line with state-of-the-art performance on similar tasks \cite{dhamani2019using,islam2020deep,luceri2020dont}. We suggest this may be due to a shift in tweet topics over time\footnote{Prior work has shown that supervised classifiers trained on Twitter data from one time period may not perform as well on Twitter data from another time period \cite{alizadeh_content-based_2020}. Tweets from the takedown dataset that were used to train the classifier were mostly from 2019 with some tweets in the beginning of 2020; tweets from the live dataset were all from 2020, with more tweets towards the end of the year due to the Twitter API's rate limits.} or a subtle adjustment of strategy by the network after the large-scale shutdown in 2020. 
To explore this further, we perform topic analysis using an implementation of BERTopic \cite{grootendorst2022bertopic}. We use DistilBERTurk embeddings, UMAP for dimensionality reduction, HDBSCAN for clustering, and \citeauthor{grootendorst2022bertopic}’s (2022) class-based variant of TF-IDF (c-TD-IDF) to get topic keywords for interpretation. We apply these methods to the takedown users’ tweets used to train, validate, and test the classifier and the live suspended users’ tweets that we evaluate on. 
Our analyses suggest topical drift between the two datsets. The top 3 topics from the takedown users’ tweets are related to finance, prison, and agriculture, whereas the top 3 topics from the live suspended users' tweets are related to society/politics, national accounts, and Turkish cities and names. 

\begin{table}[t]
\small
\begin{center}
\begin{tabular}{|l|l|l|l|l|}

\hline \textbf{ } & \textbf{Accuracy} & \textbf{Precision} & \textbf{Recall} & \textbf{F1-Score}\\ \hline

Training ($e_{1}$) & 0.734 & 0.766 & 0.574 & 0.656\\ \hline
Validation ($e_{1}$) & 0.872 & 0.885 & 0.861 & 0.873\\ \hline
Training ($e_{2}$) & 0.931 & 0.954 & 0.911 & 0.932\\ \hline
Validation ($e_{2}$) & 0.867 & 0.817 & 0.954 & 0.880\\ \hline
Testing & 0.866 & 0.814 & 0.968 & 0.885\\ \hline
Evaluation & 0.733 & 0.645 & 0.787 & 0.709\\ \hline

\end{tabular}
\end{center}
\caption{\label{tab:model_results} Results from training at epochs 1 and 2, validation at epochs 1 and 2, testing on the takedown users and evaluation on our live users that have been suspended by Twitter as of Dec 6 2021.}
\end{table}

\noindent \textbf{Experiment.} After evaluating our model on the live suspended data, we run an experiment to determine if further fine-tuning aids our performance. In February 2021, we performed the first check on our live users and found that 1,105 of our 7,973 had been suspended. This provided the first timestamp ($t_{1}$) of suspended users from the live dataset. We use this first set of suspended users at $t_{1}$ (hereafer, $S_{1}$) to train an additional layer on our classifier. After removing $S_{1}$ from the live suspended data ($t_{2}$), we are left with 1,433  users (hereafter, $S_{2}$). We collect and use an additional sample of pseudo-random Turkish language Twitter users as negatively labeled samples to create a balanced dataset for training, validation, testing on $S_{1}$ and evaluation on $S_{2}$. We follow the same BERT specifications as listed above. Table \ref{tab:exp_results} shows the results from this experiment. Further fine-tuning on our model using suspended users from the live dataset at $t_{1}$ to predict which users will be suspended at $t_{2}$ performs very well. 

Before further fine-tuning, the classifier achieves higher recall than precision, indicating that it has a higher rate of predicting false positives than of missing true positives. However, after further fine-tuning, the classifier achieves higher precision than recall. Since this task has a significant cost associated with false positives (predicting ordinary Turkish citizens as Turkish state linked accounts), a classifier with high precision is favored. The experiment shows that further fine-tuning with suspended users from $S_{1}$ results in a better-performing classifier. 

\begin{table}[h!]
\small
\begin{center}
\begin{tabular}{|l|l|l|l|l|}

\hline \textbf{ } & \textbf{Accuracy} & \textbf{Precision} & \textbf{Recall} & \textbf{F1-Score}\\ \hline

Testing & 0.834 & 0.840 & 0.801 & 0.820\\ \hline
Evaluation &  0.819 & 0.866 & 0.745 & 0.801\\ \hline

\end{tabular}
\end{center}
\caption{\label{tab:exp_results} Results from experiment using our first check of suspended users in Feb 2021 as additional training, validation, and testing data and evaluating on our live users suspended in Dec 2021.}
\end{table}


Results presented in this section provide evidence that the suspended users from our live dataset are very likely part of the AKP operation conducted on Twitter. We found that in Jan 2020 accounts in the the takedown dataset interacted substantially with live accounts through retweets and follow trains, which are common ways to boost engagement and visibility and were heavily used by the takedown accounts throughout the operation \cite{sio1}. Our BERT-based classifier achieves an F1-score of 0.71 when evaluated on the live suspended users and 0.80 after further fine tuning on our first set of suspended live users. These results indicate reasonable discernment between Turkish state-linked accounts and ordinary Turkish Twitter users, and provides evidence 
that AKP-linked IO are ongoing beyond the accounts shared by Twitter in June 2020. Following, we explore strategies of Turkish IO and provide a taxonomy of Turkish state-linked accounts.




\section{RQ2: Strategies for Resilience}

We identify strategies employed by Turkish IO actors to gain visibility and retain their social networks in anticipation of and in the aftermath of detection. Network statistics provide evidence that Turkish IO actors use explicit and implicit signals to construct their network to be robust to shutdown.

\subsection{Account Types and Role Assignment}
\label{sec:account_types}
We first identify distinct account types based on information disclosed within the user profile description. We introduce a taxonomy for the Turkish-linked accounts across two dimensions: account type and account membership. 


We define four mutually exclusive account types observed among both the takedown dataset and the live dataset: (1) \textit{main account} - the account an actor uses most often to post original content and to spread and amplify content from other accounts;
(2) \textit{retweet account} - an account that an actor uses primarily to amplify own and others' content through retweets;
(3) \textit{backup account} - an account that an actor creates in addition to their main account in case their main account is suspended and uses mostly to post retweets; 
and
(4) \textit{sequel account} - an account created by an actor in place of a previously suspended account. 

Main, retweet, and backup accounts are explicitly disclosed in two ways: (1) an account's profile description states in some form ``this is my main/retweet/backup account''; or (2) a user's main/retweet/backup account explicitly mentions (@) its backup/retweet/main account username in its profile description. 
Similarly, some accounts disclose more than one additional account in their profile description. For example, one user 
discloses their main account and two distinct retweet accounts:
\begin{quote}
\textit{Profile Description (translated): }main account - @ ibrahimk\_rael, rt 1 account - @\textit{suppressed} 
, rt 2 account @\textit{suppressed}. 
\end{quote}

\begin{figure}[tb]
\includegraphics[scale=0.3]{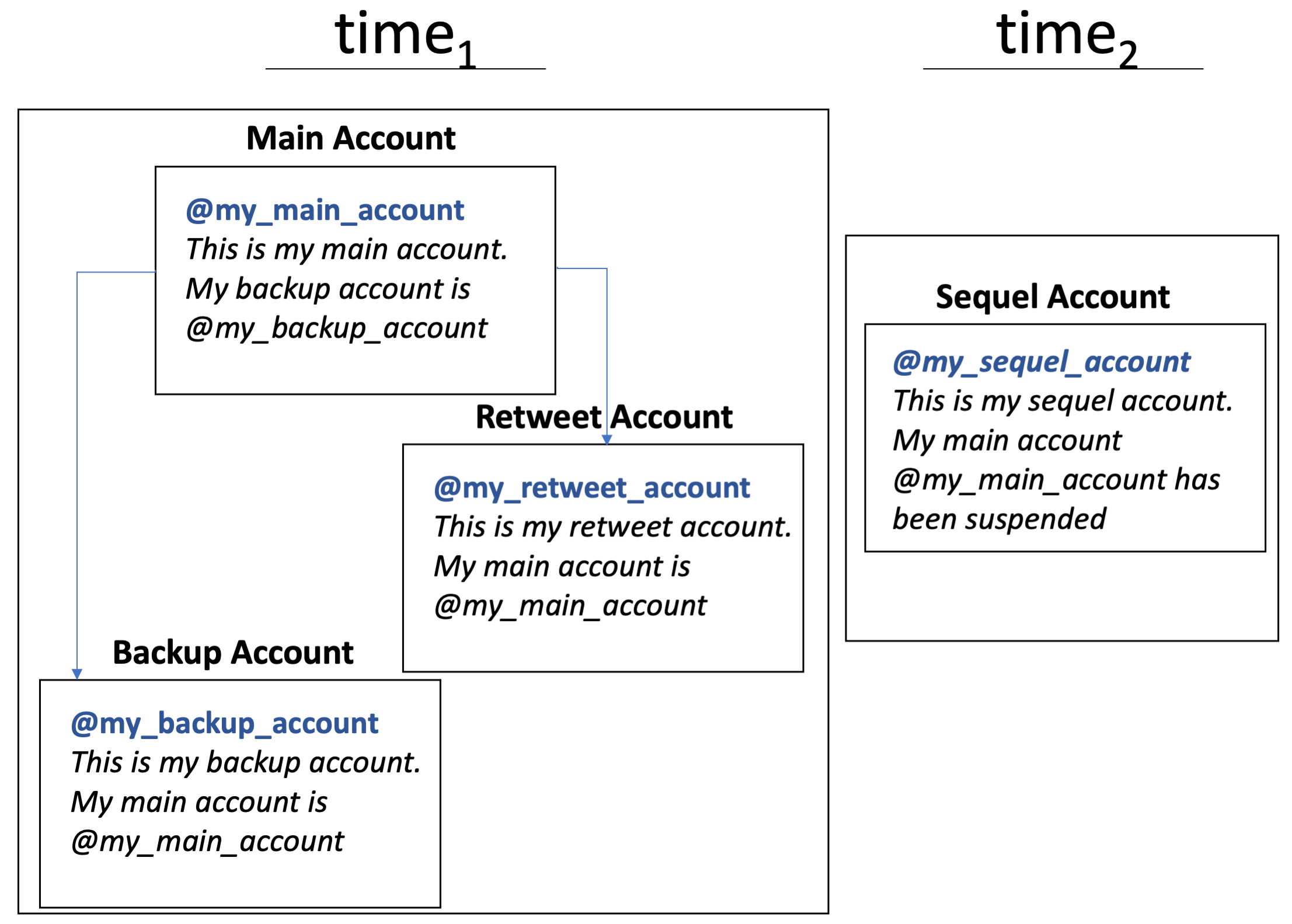}
\caption{Exemplary structure of main, retweet, backup and sequel accounts. 
}
\label{fig:account_types_figure}
\end{figure}

Main, retweet, and backup accounts are created and active during the same time period, whereas sequel accounts are created after a main account is suspended (see Figure \ref{fig:account_types_figure}). 

Table \ref{tab:acc_types_retweets} gives the number of retweets and original tweets for main, retweet, and backup accounts among both datasets.\footnote{We note that our present labelling categorizes main accounts only through explicit mentions of such in the account's profile description and therefore the group we highlight here represents a subset of such accounts.} Main accounts post the highest proportion of original content, retweet accounts post the highest proportion of retweets, and backup accounts post mostly retweets across both datasets. 

\begin{table*}[t!]
\small
\begin{centering}
\begin{tabular}{|l|l|l|l|l|l|}
\hline
\textbf{Dataset} & \textbf{Account Type} & \textbf{Total Tweets}  & \textbf{\# Retweets} &\textbf{ \# Original} &\textbf{ \% retweets}\\ \cline{1-6} 
Takedown & Main & 162,373 &110,106 & 52,267 & 67.8\\ \cline{2-6} 
& Retweet & 1,178,639 & 1,129,094  &  49,545 &  95.8\\  \cline{2-6} 
& Backup & 980,482 & 877,104  & 103,378 &  89.5\\  \cline{2-6} 
& All tweets & 36,948,536 & 27,522,156  &  9,426,380 &  74.4\\  \hline

Live & Main & 58,688 & 27,492 & 31,196 & 46.8\\ \cline{2-6} 
& Retweet & 117,251 & 111,479  &  57,72 &  95.0\\  \cline{2-6} 
& Backup & 188,744 & 162,858 & 25,886 &  86.2\\  \cline{2-6}
& All tweets & 9,298,325 & 4,890,405  &  4,407,920 &  52.6\\  \hline
\end{tabular}
\caption{\label{tab:acc_types_retweets} Original and retweet statistics for main, retweet, and backup accounts across both datasets.}
\end{centering}
\end{table*}



 \begin{table}[t!]
 \small
 \begin{centering}
 \begin{tabular}{|p{0.95\linewidth}|}
  \hline
 \textbf{Profile Description (translated)} \\ 
 \hline 
My old account has been suspended, this is my new account. \\ \hline
Bastard feto guys, you close it, I will make a new account \\ \hline
New account. I came back. You won't be able to silence it. \\ \hline
New account Only national accounts can be followed immediately. (Either you rise with Islam, or you rot with denial ...) \\ \hline
new account, my other account is CLOSED \#BACKUPACCOUNT @\textit{suppressed} \\ \hline 
\end{tabular}
\caption{\label{tab:suspended_examples} Sequel account examples. }
\end{centering}
\end{table}

Table \ref{tab:suspended_examples} gives examples of newly created sequel accounts disclosing that their previous accounts have been suspended. 
We hypothesized that the live dataset may contain direct sequel accounts from the takedown dataset. Following, we describe this analysis. 

\subsection{Direct Sequels Between Datasets}
\label{sec:direct_sequels}
We aimed to identify whether a subset of users in the live dataset are sequels for accounts in the takedown dataset. 
We calculated similarity scores for usernames, profile descriptions, and display names between the two datasets. First, we lowercased all usernames and calculated the Levenshtein distance ratio between each username in both datasets. For takedown users, we match the user from the live dataset with the maximum Levenshtein distance ratio, which we approximate as the user who is most likely a sequel account. Once we have these user pairs, we calculate the similarity between each pair's profile descriptions and display names using the Python library difflib's SequenceMatcher.ratio\footnote{This returns any hashable sequence's similarity as a float between 0 and 1; 1 if the sequences are identical, and 0 if they have nothing in common.} 
method. We also calculate the number of 3rd party users that both accounts in the user pair interacted with under the assumption that sequel accounts will retweet, reply to, quote, and mention some of the same users. 
We set a maximum Levenshtein distance ratio $>$ 0.9 for usernames, \textit{or}
a maximum Levenshtein distance ratio $>$ 0.6 for usernames \textit{and} at least one of the following: a similarity score $>$ 0.5 for all non-null profile descriptions in the user pairs, at least 2 common interactions between the user pairs, or display name similarity score $>$ 0.8. This gave us 169 user pairs between the takedown dataset and the live dataset that we classify as sequel accounts. Table \ref{tab:username_similarity} gives examples of sequel pairs.


 \begin{table*}[h!]
 \small
 \begin{centering}
\begin{tabular}{|p{0.13\linewidth}|p{0.13\linewidth}|p{0.13\linewidth}|p{0.15\linewidth}|p{0.13\linewidth}|p{0.13\linewidth}|}
\hline
\textbf{Takedown} & \textbf{Live} & \textbf{Username Similarity} & \textbf{Bio Similarity} & \textbf{Name Similarity} & \textbf{\# Common Interactions} \\ \hline
\textit{suppressed} & \textit{suppressed} & 0.963 & 0.327 & 0.231 & 0 \\ \hline
\textit{suppressed} & \textit{suppressed} & 0.933 & 0.254 & 0.920 & 0 \\ \hline
\textit{suppressed} & \textit{suppressed} & 0.929 & 0.680 & 1.000 & 0 \\ \hline
\textit{suppressed} & \textit{suppressed} & 0.897 & 0.328 & 1.000 & 0 \\ \hline
\textit{suppressed} & \textit{suppressed} & 0.897 & 1.000 & 0.919 & 2 \\ \hline
ihsantopbas & ihsan\_topbas42 & 0.880 & 0.114 & 0.812 & 5 \\ \hline
avhasanteke & av\_hasanteke27 & 0.880 & 0.526 & 1.000 & 16 \\ \hline
hocaketum & hocaket & 0.875 & 0.125 & 0.930 & 11 \\ \hline
\textit{suppressed} & \textit{suppressed} & 0.783 & 0.108 & 0.875 & 1 \\ \hline
\end{tabular}
\caption{\label{tab:username_similarity} User pairs that we classify as sequel accounts given their username similarity, user profile description similarity, user display name similarity, and the number of common users with which that they interact. \emph{Usernames of users with fewer than 5,000 followers are suppressed for privacy.}}
\end{centering}
\end{table*}

\subsection{Account Membership}

In contrast to accounts associated with other IO, e.g., those originating in Russia, Saudia Arabia, and China, which are believed to typically to hide their affiliation with the state \cite{arif_acting_2018, king2017chinese, sio2}, 
many AKP-linked accounts explicitly mention their allegiance to AKP groups and their allegiance to the state through national accounts and group accounts. 

We define \emph{national accounts} as those that explicitly mention ``this is a national account'' or use any of the hashtags \#MilliTakipMerkezi (\#NationalFollowingCenter), \#MilliHesaplarYanyana (\#NationalAccountsSideBySide), and \#MilliHesaplarBurada (\#NationalAccountsHere) (see Table \ref{tab:acc_tyes}).
National accounts disclose in their tweets and profile descriptions that they are building an alliance of followers online in support of the Turkish government, specifically the AKP. We find that the 
these accounts encourage follow trains to increase visibility. For example:
\begin{quote}
\textit{Translated tweet from @\textit{suppressed} (user from the live dataset)}: 
FRIENDS, DON'T HAVE A SMALL ACCOUNT!!
In order to strengthen national accounts,
We support the work of
\#NationalAccountsTogether, which started with the instruction of our President
\#NationalAccountsHere RT and Let those who reply follow each other.
\end{quote}

National accounts' hashtags are some of the most tweeted hashtags among our live tweet data. \#NationalAccountsSideBySide (tweeted 54,526 times) is the \#1 most tweeted hashtag, followed by \#NationalAccountsHere (tweeted 27,835 times). While, \#NationalFollowingCenter was tweeted 7,428 times. We suggest that the IO is using national accounts to encourage ordinary citizens to engage with their content and begin posting their own pro-government content.
 \
\begin{table*}[t!]
\small
\begin{center}
\begin{tabular}{|l|l|l|p{0.6\linewidth}|}

\hline \textbf{Account Type} & \textbf{Live} & \textbf{Takedown} & \textbf{Profile Description (translated)}\\ \hline

Main & 34 & 23 & My main account. RT Account: @ibrahimergin98 \\ \hline
Retweet & 59 & 92 & RT ACCOUNT. Let the storms stop, give way to the Turkish flag.\\ \hline
Backup & 112 & 68 & BACKUP ACCOUNT. MAIN ACCOUNT: @\textit{suppressed}.\\ \hline 
Sequel & 47 & 27 & New account, old one is suspended!\\ \hline
National accounts & 214 & 6 & THE REPUBLIC ALLIANCE IS FOLLOWING ACCOUNTS. \#NationalFollowingCenter. \#NationalAccountsTogether.  \#NationalAccountsHere.\\ \hline
Enderun group & 10 & 10 & ENDERUN 5 RT \& FAV accounts. Please write DM to join our \# EnderunGroups. Founding President @\textit{suppressed}\\ \hline 

\hline
\end{tabular}
\end{center}
\caption{Number of users and examples by type.}
\label{tab:acc_tyes} 
\end{table*}

We define \emph{group accounts} as those which explicitly mention their membership in an AK Twitter group. In their report on the Turkish accounts, Stanford Internet Observatory (SIO) refers to these groups as ``retweet rings,'' because of their highly organized and centralized pro-AKP retweet structure \cite{sio1}. We identified nine distinct groups within our live dataset that follow the same structure as the retweet rings described in SIO's report. This structure includes nearly identical profile descriptions declaring group membership, and designated group emojis in the user display names and profile descriptions. For example, accounts belonging to ReisiOsmanlıGrupları all use the same yellow star emoji on both sides of the group name, accounts in EnderunGrupları use the red ``anger emoji'', and accounts in REİS61 use the Turkish flag. The Enderun group is present in both the takedown dataset and the live dataset. Table \ref{tab:acc_tyes} gives the number of samples of each account type and membership across takedown and live datasets.

As indicated by the presence and coordination of backup and sequel accounts, AKP-linked IO actors anticipate shutdown and take measures to continue their agenda after account suspension. We hypothesize that the network may use explicit signals strategically to garner easy visibility while depending on more subtle signals to provide a social network backbone even in the face of shutdown. We explore this in the following section.

\subsection{Network Experiment: Explicit v.s. Implicit Nodes}
\begin{table*}[!h]
\small
\begin{centering}
\begin{tabular}{|p{0.17\linewidth} | p{0.10\linewidth} | p{0.10\linewidth} | p{0.08\linewidth}|p{0.10\linewidth}|p{0.10\linewidth}|}
\hline
\textbf{Subnetwork} & \textbf{Nodes} & \textbf{Edges} & \textbf{Density} & \textbf{Diameter} & \textbf{Path Length} \\ \hline
Full graph & 11,551 & 609,459 & 0.005 & 15 & 3.525 \\ \hline
Implicit only  (Explicit removed)  & 10,635 & 474,833 & 0.004  & 16 & 3.627 \\ \hline
Explicit only  (Implicit removed)  & 915 & 9,877 & 0.012  & 8 & 3.339 \\ \hline
Random implicit only & 10,636 & \textit{varies} & 0.005 & 15.2 &  3.536\\ \hline
Random explicit only & 915 & \textit{varies} & 0.005 & 11 &  3.573\\ \hline
\end{tabular}
\caption{\label{tab:easy_hard_stats} Graph statistics (density, diameter, avg. path length) for the network in Figure \ref{fig:all_interaction_graph} and subnetworks corresponding to: all nodes, implicit nodes, explicit nodes, random samples for comparison, nodes in the takedown dataset only, nodes in the live dataset only, and explicit and implicit nodes in the takedown dataset and the live dataset only. Number of edges for random samples varies with each sample.}
\end{centering}
\end{table*}

In 2020, Twitter removed thousands of Turkish IO accounts, yet, our findings show that the operation has maintained a robust presence on Twitter. After identifying explicit traces of IO strategy, such as dedicated backup and retweet accounts, and direct sequels for accounts taken down by Twitter, we hypothesize that Turkish IO actors utilize explicit signals to support engagement with 
their followers. While, accounts with no explicit signals are purposefully kept discreet to avoid detection and serve as a structural backbone of the operation. 
We test these hypotheses by way of graph statistics for the network in Figure \ref{fig:all_interaction_graph} and various subnetworks.

We classify \emph{explicit nodes} as accounts that are: (1) main, (2) retweet, (3) backup, (4) sequel (including direct sequels), (5) group, or (6) national accounts, or if they mention ``add to groups'' or ``do not add to groups'' in their profile description. All others are considered \emph{implicit nodes}. In total, we have 915 explicit and 10,636 implicit nodes.

We calculate network density, diameter, and average path length over five distinct subnetworks from Figure \ref{fig:all_interaction_graph}. 
Our hypothesis suggests that the network structure should remain relatively unchanged when removing explicit nodes. 
While, if we remove all implicit nodes, we expect graph metrics to significantly change. In this case, we are removing a large majority of the network, and in particular, nodes we suggest are integral to the longevity of the network. 

Finally, we also remove five random sets of 915 nodes and five random sets of 10,636 nodes as ``random explicit'' and ``random implicit'' baselines, respectively. 
These random samples follow the same class distribution (takedown vs. live) as their comparable subnetworks. 

Table \ref{tab:easy_hard_stats} details the results of this experiment. Graph density decreases by 20\%, diameter increases by 1 (6.67\%), and the average path length increases by 2.8\% after removing the explicit nodes. After removing the implicit nodes, the graph density increases by 140\%, the diameter decreases by 46.7\%, and the average path length decreases by 5.3\%. However, we do not find significant changes after removing equally-sized random samples of nodes. These results reveal that the subnetwork of \textit{only} explicit nodes is a denser, more closely connected network. Whereas, the network of \textit{only} implicit nodes has similar properties to the full graph. 
This supports our hypothesis that the operation utilizes explicit signals to gain visibility and retain a loyal set of followers but is likewise supported by a robust subnetwork of discrete accounts structured to withstand detection and shutdown. 



\subsection{Generalizability to Other Countries}
We conduct parallel analyses on explicit v.s. implicit signals for archived IO originating from Russia and China to examine if our findings from the Turkish network generalize to other countries. To do so, we use Twitter's Oct 2018 and Jan 2019 releases of Russian/IRA accounts (here forward, R1 and R2, respectively) and Sept 2019 and May 2020 releases of Chinese accounts (here forward, C1 and C2, respectively).\footnote{As with the Turkish dataset, these accounts and their activities can be downloaded on Twitter's Elections Integrity Hub in the Information Operations archives. https://transparency.twitter.com/en/reports/information-operations.html\#1.3 } R1 contains 3613 accounts, R2 contains 416 accounts, C1 contains 4301 accounts, and C2 contains 23,750. We note that, in the case of the Turkish IO, there has been just a single release of state-linked IO shared by Twitter so our collection of live accounts serves as a second time point to explore the ongoing operation. In the cases of Russia and China, Twitter has shared multiple releases of IO over time thus supporting a similar inquiry without live account collection.

We first run the direct sequels analysis on both networks of accounts and find 151 direct sequels between C1 and C2, but only 3 direct sequels between R1 and R2. Next, we aim to find additional explicit signals from the Chinese and Russian operations. 
We manually label all Russian and Chinese IO for presence of explicit signals and sketch primary findings. We find that while the Russian and Chinese accounts do not exhibit the same explicit signals as the Turkish network, they do exhibit their own, and we find interesting overlap between the two groups' explicit signals. 

\noindent \textbf{Russian IO.} We observe five primary categories of explicit signals within the Russian network: (1) local news accounts for US cities (e.g., ``Breaking news, weather, traffic and more for St. Louis. DM us anytime. RTs not endorsements''); (2) fake personas portraying black Americans and southern conservative Americans or organizations (first identified in \cite{arif_acting_2018}, e.g.: ``Slave blood. Melanin. Water. Fire. Honey \#BlackLivesMatter bih,'' ``Christian Conservative. \#Trump Supporter \#Election2016. Watching the polls. Make America Great Again  \#WakeUpAmerica \#tcot''); (3) accounts containing a list of repeated buzzword characteristics with ``USA'' as the location (e.g., ``Lifelong food aficionado. Hardcore problem solver. Twitter lover. Typical bacon fanatic. Creator.''); (4) accounts with inspirational/motivational/love sayings in the profile description and ``USA'' as the location (e.g., ``It's not about being the best. It's about being better than you were yesterday''); and, (5) personas of Russian immigrants living in the US (e.g., ``Russian Immigrant currently living in the US''). We find 647 explicit accounts within the Russian IO datasets.

\noindent \textbf{Chinese IO. } The Chinese network displays four distinct categories of explicit signals, with some overlap with the Russian network: (1) accounts that list followback hashtags in the profile description (e.g., \#autofollowback, \#teamfollowback); (2) accounts that have lowercased keyboard mashing in the profile description (e.g., ``hgfdsa,'' ``vdcwqd''); (3) accounts that contain love and marriage sayings in the profile description with location set to a US city (e.g., ``A great marriage is not when the ‘perfect couple’ comes together. It is when an imperfect couple learns to enjoy their differences.'' Location: ``Knoxville, USA''); and, (4) accounts containing a list of repeated buzzword characteristics (e.g., ``Food aficionado. Alcohol expert. Introvert. Proud tv maven. Award-winning zombieaholic. Wannabe coffee scholar. Music guru. Thinker.''). The categories containing lists of buzzword characteristics in the Russian and Chinese networks have significant overlap and both use words such as aficionado, expert, lover, junkie, trailblazer, maven, guru, and more, following items such as coffee, alcohol, beer, social media, music, and more. We find 862 explicit accounts within the Chinese IO datasets.

\noindent \textbf{Comparative analysis.} The explicit signals identified within the Russian and Chinese networks do not show similarity with explicit signals identified in the Turkish network aside from one interesting overlap. We identify two backup accounts for the Russian account ``@TEN\_GOP.'' @TEN\_GOP was an IRA-run fake Tennessee GOP account that amassed over 150,000 followers, including Donald Trump, Jr. This account was highly significant in the IRA's campaign targeting the 2016 US Presidential Election, and has been the subject of multiple news articles \cite{mueller,timberg_dwoskin_entous_2017, vox,cnn,tennessean}.

We create account interaction graphs for both countries and conduct a network analysis identical to the experiment shown in Table \ref{tab:easy_hard_stats} for Russia and China's networks. Network visualizations and experiment results are included in \textit{Supplemental Materials}.\footnote{https://github.com/mayamerhi/Turkey\_IO\_ICWSM23} We do not find significant increase in density and decrease in diameter and average path length when examining the explicit only vs. the random explicit only subnetworks as seen in the Turkish network. 

From these findings, we conclude that the presence of explicit signals in IO accounts generalizes across the three countries we have examined. Yet, we find that the nature of explicit signals are largely unique to each country's operation, aside from the overlap in buzzword characteristics used in both the Russian and Chinese networks. We find that the account types and group memberships we present in this work are, to our knowledge, unique to AKP-linked accounts. We also show that our methodology for identifying direct sequel accounts was successful in finding a significant number of direct sequels in the Chinese network, but not in the Russian network.

Explicit signals identified in the Russian and Chinese networks are individual in nature, whereas the group and national accounts from the Turkish network have a group structure. These explicit group structures are, to our knowledge, unique to Turkey. We suggest that the dense explicit only subnetwork (shown in Table \ref{tab:easy_hard_stats}) is likely due to this group structure.
Our findings highlight the lack of a one-size-fits-all solution for studying and mitigating IO. 

\section{Discussion and Conclusions}
\label{discussion}

Turkey has been known to operate state-linked Twitter accounts since at least 2013 using a grassroots approach, including hiring young tech-savy Turkish citizens to create and disseminate pro-government content on social media \cite{albayrak_turkeys_2013, saka_social_2018}. We present 7973 Twitter accounts collected from 13 parent accounts that interacted with and resembled AKP-linked accounts from Twitter's takedown. The accounts we present here are a subset of what is likely an expansive and dynamic network of Turkish IO accounts. Our subset is dependent upon the initial parent accounts; had we chosen different parent accounts, our resulting set of accounts would have been different. Of these accounts, 2454 have been suspended since collection. Due to their grassroots approach, it is incredibly difficult to distinguish Turkish IO actors from ordinary AKP supporters. We have not studied whether there are ordinary AKP supporters in our live dataset, and we do not claim that every account in our dataset is linked to the operation. However, given Twitter's subsequent suspension of 2454 of our live accounts, the reasons Twitter states for suspending users, and the results of our classifier, we believe that this subset of our live dataset is likely linked to Turkey's ongoing IO.  



We utilize machine learning techniques to show latent similarities between content from Twitter's takedown dataset and our live dataset. We identify strategies that Turkish IO actors appear to utilize to remain resilient to shutdown. These include explicit signals to bolster alliance and visibility, such as national account hashtags and groups, which we suspect are also used to engage ordinary citizens and appear as grassroots government support.
After accounts are shutdown, IO actors preserve their social networks by standing up highly similar replacement accounts which are easy for their followers to find, even if it makes them more vulnerable to detection again. Our findings suggest that the integrity of the operation relies on a set of discreet accounts to keep the network intact as explicit accounts are detected and shut down. We believe that there is likely also more subtle strategy supplementing operations manifest as implicit relations amongst accounts that may be tied to tweet text and media. Next steps in understanding the strategies employed by the Turkish IO campaign should explore this content and potential shifts in related tactics, e.g., use of memes, cross-platform engagement. 

To examine whether our findings in the context of Turkish operations generalize to our countries, we apply similar analyses to two sets of Russian and Chinese IO accounts archived by Twitter. Our methodology for identifying direct sequels generalizes well to China, but not to Russia. Both Russia and China exhibit a robust set of explicit signals, however, they are unique to each country for the most part. Results of comparative network experiments suggest that Turkey uniquely constructs a dense subnetwork of explicit accounts. We attribute this to Turkey's unique group structure discussed in the Account Membership analysis.


We have suggested that Turkish IO actors began anticipating and preparing for shutdown by opening backup and retweet accounts, and after shutdown, sequel accounts. This is a novel finding that we did not anticipate at the time of data collection when we decided to limit account collection to those created in 2020, which led to a very small window of activity overlap between the two datasets studied here. This is a limitation of data collection and we suggest future work studying IO dynamics include accounts created prior to large scale shutdown. 


Our findings highlight the time, effort and cost imbalances between IO actors and the institutions working to combat them. For IO actors, opening a large number of social media accounts is relatively cheap and easy. For Twitter, significant resources are expended to detect them and account shutdown bears risks. Broadly, it seems that Twitter's efforts to mitigate coordinated manipulation on their platform is a step in the right direction for creating an open, democratic information environment. Their data sharing provides researchers with exceptional volumes of ground-truth IO data and they set valuable industry standards for transparency. However, our findings also suggest that current approaches fall short and highlight the need for fundamental shifts in platform norms and policies. 

\section{Ethics}
\label{ethics}


This work provides novel and valuable insights to the research community studying IO and coordinated influence online. We provide our dataset of a suspected live Turkish IO on Twitter, along with our random sample of Turkish-language Twitter users used to train our classifier. Both datasets are available on the first author's Github page.~\footnote{https://github.com/mayamerhi/Turkey\_IO\_ICWSM23} Following Twitter's Terms of Service, we release only the User ID's and Tweet ID's for all data presented here ~\footnote{https://developer.twitter.com/en/developer-terms/policy}. Also following Twitter's protocols, we suppress the usernames of all users with fewer than 5000 followers.

In this work, we also obtained and analyzed Twitter's unhashed versions of their archived IO data. Academic researchers must apply for access to the unhashed versions of the data, provide details on the proposed usage, analysis, storage of the data, and sharing of results obtained from analyzing the data, and sign a data agreement. The first author obtained access to the unhashed data, and followed all guidelines laid out by Twitter. 

We expect that these datasets may be utilized to build models to detect future Turkish IO as their strategies continue to evolve. We acknowledge that some users in our live collected dataset may not be affiliated with the Turkish state and may not be attempting to influence online users through coordinated operations. We present the live dataset as suspected IO based on our observations and analyses, but we cannot assess the origins and intentions of these users with certainty given the information we have. This is true for Twitter's archive data as well, which Twitter acknowledges as do we, and which motives the suppression of usernames for users with fewer than 5000 followers.

Both datasets were collected by the first author through their personal Twitter account using the Academic Track Twitter API and stored on a password protected external hard drive. The data were also uploaded to a cloud server for analysis, which is protected with two-factor authentication. After analysis, data were removed from the cloud server and stored only on the password protected external hard drive. 

\section{Acknowledgement}
This work was in part supported by NSF awards \#1820609 and \#2131144.

\bibliography{bib}
\end{document}